# ANALYSING THE POTENTIAL OF BLE TO SUPPORT DYNAMIC BROADCASTING SCENARIOS

Miran Borić[1] Ana Fernández Vilas[2] and Rebeca P. Díaz Redondo[3]

[123]Information & Computing Lab., AtlantTIC Research Centre,
School of Telecommunications Engineering,
University of Vigo, 36210, Vigo(Spain)

## ABSTRACT

*In this paper, we present a novel approach for broadcasting information based on a Bluetooth Low Energy (BLE) ibeacon technology. We propose a dynamic method that uses a combination of Wi-Fi and BLE technology where every technology plays a part in a user discovery and broadcasting process. In such system, a specific ibeacon device broadcasts the information when a user is in proximity. Using experiments, we conduct a scenario where the system discovers users, disseminates information, and later we use collected data to examine the system performance and capability. The results show that our proposed approach has a promising potential to become a powerful tool in the discovery and broadcasting concept that can be easily implemented and used in business environments.*

## 1. INTRODUCTION

The main characteristics of a smart environments is their adaptability and awareness of their surroundings. Such adaptable ability provides the Internet of Things concept by its linking the virtual with the physical world. This whole process presents a challenge since the smart environment is heterogeneous and supported by many different technologies[1][2]. One of these technologies on the rise is a BLE beacon, which found a place in many aspects of human needs and living. Such technique enables smartphones, tablets, and different wearables to perform a particular kind of action when near a beacon. Nowadays, many companies are involved in the beacon production[3][4]. However, the price is controlling the beacon solution, especially if it supports some enhanced capabilities. Also, many companies with the hardware part include their application solution. That is to say; the application often comes incombination with the hardware they offer. Because of that, beacons are not usually capable of analysing the surrounding around them (user discovery) without some previously installed app on a user's smartphone[5].

This paper presents a fully customized solution for user discovery and dynamic beacon information broadcasting according to the discovered user. In our case, a single beacon device is capable of supporting a dynamic broadcasting by predefining user profiles through the use of some background intelligence. Therefore, without any extra equipment, a single beacon device can serve different profiles at the same moment. Also, the possibility of monitoring user's movements, without any additional proprietary app installation, gives us an opportunity to gather more information about the space analytics and user behaviour.

This solution can be used in any message dissemination environment and integrated into the existing infrastructure with minimum investments. In our case, we have tested it as an indoor solution for broadcasting faculty information to students, faculty visitors, and faculty personnel. We verify the system through the experiment in a particular period and obtain the data. Later, by analysing the data, we extract and examine the results by giving a better insight of system performance and system usage. Furthermore, we investigate the capability of the system to receive interactions during the experiment period and how these interactions can show different activities in the area.

We divide this paper into several sections. Section 2 gives the related work and what motivated us for this paper. In section 3 we give the system explanation and the architecture. Section 4 deals with pilot experiment and shows us all the details regarding the experimented. The last part draws the conclusions, current and future work.

## 2. RELATED WORK

In [6]authors offered a crowd sensing solution where users receive a corresponding information by wearing a Bluetooth bracelet, which connects to their smartphones. Furthermore, specific bracelets connect to the beacon devices placed along the experiment area. Also, a dedicated server controls all the beacon devices and that way it manages a user-bracelet-beacon interaction.However, in our case, the interaction is almost similar but without Bluetooth bracelets. Our system makes a direct connection between beacon advertising device and a user's smartphone, which eases the whole process.

The most significant influence and motivator of our work is a commercial solution offered by Cisco Systems, Inc. called Cisco Connected Mobile Experience (CMX)[7]. This product presents a solution that uses location intelligence gathered from the Wi-Fi network to help companies of all various industries monetize their network by engaging with their clients and optimizing marketing operations. It offers to businesses a unique way to communicate with and to better understand their customers while improving customer experience and their venue. CMX system can automatically detect the presence and engage their customers on their site and with tailored messages. Based on a user's position and by using his smart mobile device, the system can deliver different content. Ranging from informational updated, indoor maps or any user relevant information this location data also increases the efficiency by understanding new versus repeat customers.

Besides being a powerful mobile solution the price can be its disadvantage. Not all organizations are ready to invest in such solution. Implementing it usually requires a Cisco proprietary devices and its core, e.g., Cisco 3365 Mobility Services Engine. In our case, we offer an inexpensive system installed on a Raspberry Pi device that can discover users and distribute them the corresponding information to improve organizational and operational efficiency. Also, it can understand a user pattern, send different data to identified users, etc.

## 3. ARCHITECTURE OVERVIEW

The reason we choose BLE beacon technology is the fact that presents a lightweight protocol that does not ask for prior connection between two devices that need to share some data. BLE beacon-enabled device broadcasts the signal to anyone who is configured to listen. In case of the Wi-Fi direct and classic Bluetooth, they ask for secure pairing to unknown devices[8]. Furthermore, if we want to connect multiple clients to Wi-Fi network, devices also require a specific data routing protocol that is self-organized[9].A beacon represents a BLE class, and sometimes it is called a BLE flavor. It became popular thanks to its Bluetooth low power capability to stay active for years and it is powered only by coin cell battery[10]. Among many beacon solutions[11], for our system implementation, we use the Apple version (ibeacon), which we found most suitable for our pilot project. The technology is pretty much straightforward to explain. It represents a transmit-only technology built for broadcasting data, where receiving option demands additional configuration into the beacon device. However, the protocol itself is transmit-only. Put it more understandable, ibeacon, while advertising packets, it is doing nothing more than saying to everyone around that it is present and identified by three numerical identifiers: Universal Unique Identifier (UUID), Major number and Minor number. Combining these three numbers we can representa specific information, and by advertising this combined identifier, it can trigger a particular action on a device that reads it. What is interesting about the ibeacon is the application part, usually installed on a user's phone.Such app can read the ibeacon identifier and associate it with the action that starts after discovering the beacon. In this manner, an application function on a user's phone is the primary product throughout the whole beacon process, whether it is designed to show some information regarding advertised products in the beacon proximity or something else more complicated.

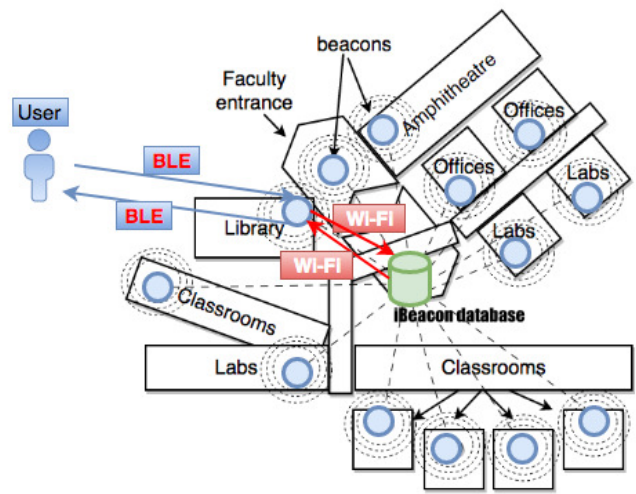

Figure 1. System architecture

We implement and test the solution at the Faculty of Engineering and Telecommunications of Vigo in Spain. We divide its infrastructure into two main parts, which are the front-end (BLE part) and the back-end (Wi-Fi part). The front-end is taking care of the user discovery and sending multiple user info advertisements, while the back-end connects all the beacon devices through the Faculty Wi-Fi infrastructure to the ibeacon user profile database (see Fig. 1). Single user, in other words, a user's device, is discovered when enters the beaconing area. Beaconing area contains several Raspberry Pi beacon configured devices places along the Faculty area. BLE Module on a single Raspberry Pi device is responsible for the user discovery process and is continually scanning for new MAC addresses over a BLE protocol. When the beacon device discovers a user, it learns a user's device MAC address and saves it to the ibeacon database.

For this process, we are using two technologies. First, a user is discovered through the BLE protocol and sent to the database by using a Wi-Fi protocol. Later, when ibeacon database saves an input for a specific user, it associates the user with a beacon configuration advertisement (UUID, Major and Minor combination) and again through the Wi-Fi protocol sends a configuration advertisement to the beacon device. From that point, a beacon device uses BLE protocol to deliver configuration advertisements to a specific discovered user. User's smartphone reads the configuration advertisement and triggers a particular action (see Fig. 1). Moreover, ibeacon database saves all the user's input records, and it is updated every time a user enters the area. It does this by keeping track of a user's beacon entrance timestamps. User discovery process is customizable and configured according to the needs of the organization. In our case, a single beacon device works in a discovery mode for five seconds, and after one second it restarts again. We have to mention that when a beacon device discovers a particular user's device, it immediately makes a connection to the ibeacon database and saves the user's MAC address as the database input. It also saves the timestamp of the user's entrance to the broadcasting (beaconing) area.

Along the Faculty area, several beacon advertisers were implemented and continuously served users while entering and exiting the Faculty area. For beacon devices, as already mentioned, we use a Raspberry Pi hardware devices[12]. The reason why we choose Raspberry Pi over many others, ready to implement commercial solutions, is the fact that they lack in customization abilities, where Raspberry is Linux based[13], and it is open to customization. Hardware part consists of few components. The primary element is a Raspberry Pi 3, and it represents a beacon device, but the ibeacon database as well.Every beacon device has two pieces of Bluetooth CSR[1] 4.0 USB[2] dongles and one Wi-Fi USB dongle installed on its chassis. The reason why we need two BLE dongles is that of discovery and advertising process. Beacon device is continuously scanning the beacon area on one BLE dongle why the other has a function of the advertising process. Wi-Fi dongle connects the beacon to the internal Faculty infrastructure, and that way communicates with the ibeacon database. For the ibeacon database, we use Raspberry Pi 3 model B with MySQL version 5.5 but any computer that supports MySQL installation will serve the purpose.

## 4. PILOT EXPERIMENT

Faculty as a building presents a complex environment, where we had to take care of the BLE coverage, which reduces if the signal encounters an obstacle (walls, doors, inventory, etc.). After positioning all the beacon devices and making sure that we conceived a full coverage, we connected them all through the Faculty Wi-Fi infrastructure. The main idea was also to build a system and satisfy the broadcasting coverage area by using a minimum number of devices. In our case, we conceived it with 13 devices.We evaluate the performance of the proposed system by collecting and extracting real-time data. We experiment on the premises of the Faculty and in 27 days, from 05/05/2017 to 31/05/2017, when the Faculty as an institution was in its usual working routine process (professors dealing with students, students attending classes, exams, studying in the Faculty library, etc.). After collecting and extracting all the data from the ibeacon database, we discovered that the total number of inputs in the database was 70,046. This info also means that the interactions of all the identified users and the system were the same as the number of database inputs previously mentioned. By interaction, we express discovery and broadcasting process altogether. Also, a total number of discovered and served users during the experiment period was 120. It is important to mention that no influence on users was done, regarding their smart devices and activating the BLE protocol, whatsoever.

Next, we examine a basic system usage from several points of view and divide it by days and hours. Usage tells us at what specific time was the system involved in a message dissemination process, as well as in the user discovery process. From Fig. 2 we can see that the peak time of system usage was 15th of May with 8,884 total interactions. Some predictions regarding such system behavior could be that the number of Faculty visitors was the same, but not all users had BLE protocol activated on their phones. However, taking into account that the exam period at the Faculty started on 15th of May, we can conclude that starting from that day, a Faculty was less crowded place than before. The reason for this is because not all students were taking the exams, and by that, they were not present like in the time when they were attending their classes (see Fig. 2). More in-depth inspection of user system usage will give us a better insight of the day with the high user interactions.

---

[1]Cambridge Silicon Radio
[2]Universal Serial Bus

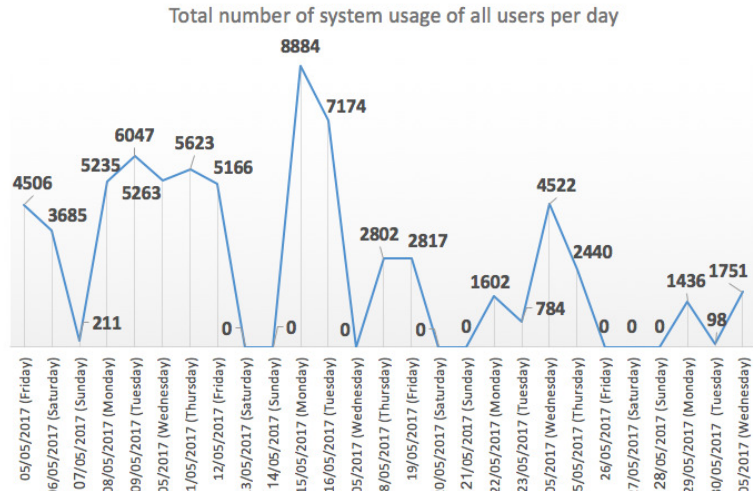

Figure 2. System usage per days

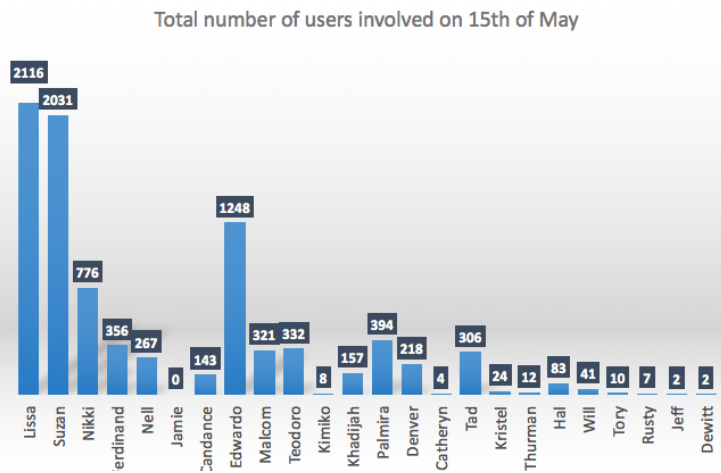

Figure 3. System usage – 15th of May

If we dive a little bit more into this specific date (15th of May), we see that the total number of users involved in system usage was 24 (see Fig. 3). Figure 3 gives us a better insight into every user interaction. More specifically, two users, named Lissa and Suzan, were pretty much involved in system interaction, comparing to the others. In the following part, we will give a closer insight of the most active user and its behavior during the experiment. It is important to mention that every user identification, in our case a user's device MAC address, we changed for username, for the sake of user's anonymity. In a real-world scenario, every username presents a specific and unique device MAC address, as well as user's device identification (phone name). Next, we wanted to know the hour peak time of system usage. According to Fig. 4 and the closer insight, we can tell that the users mostly interact from 10 AM until 12 PM. Peak time also shows that the Faculty as the institution had usual working hours conventional to the world's working hours. We can see when the day goes down, in the same manner, the system usage slowly drops to the point of around 850 interactions per hour until the next morning at about 8 AM. This information tells us that the users used system resources also during non-working hours. Why users use the system during night time is the fact that Faculty has employees who leave their electronic equipment (device's BLE protocol) in active mode. In the following paragraphs, we examine more closely involved users in this system usage process.

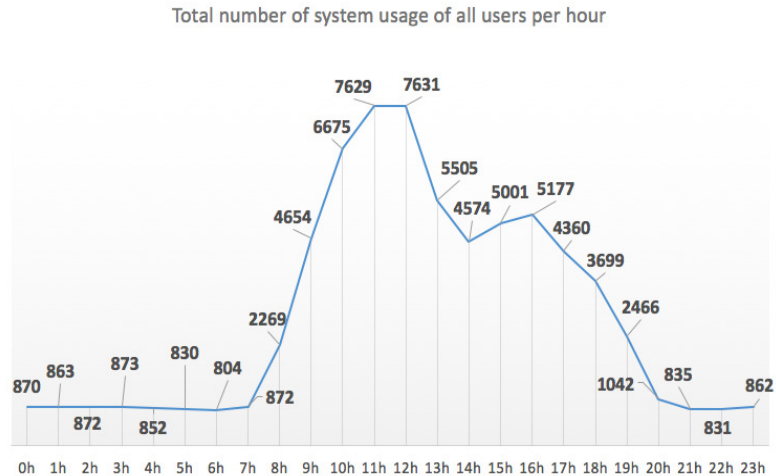

Figure 4. System usage per hour

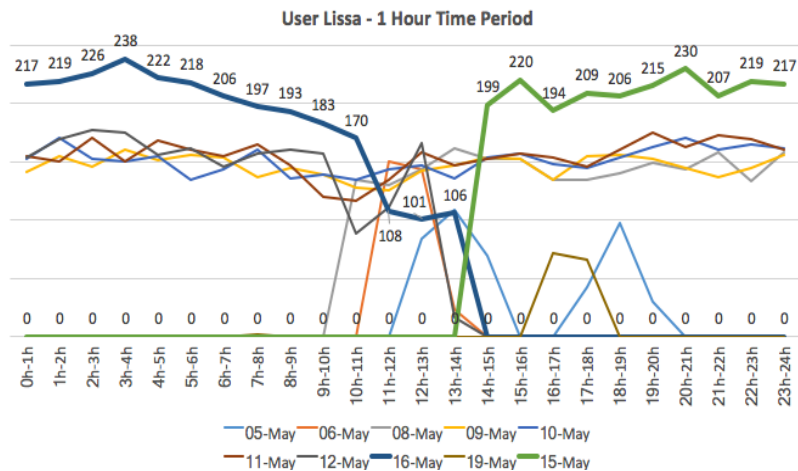

Figure 5. User Lissa interactions with the system

The user that we examined was user Lissa. This user had the most number of interactions with the system (20,476) during the experiment period of 27 days. From Figure 5 we notice that user was active from 5th of May until 19th of May and usually was leaving his BLE device turned on, even during night time when the Faculty does not operate for the public. This behavior we notice on 8th, 9th, 10th, 11th, 12th 15th and 16th of May. Notably, this action we see on 15th and 16th of May when user's device was mostly active during night period (see Fig. 5). This why we had deviation on these days (see Fig. 2). During different periods of time the user was acting according to the behavior of other scanned users, either not present in the area (out of the Faculty), or either present but with inactive BLE protocol on his device.

# 5. CONCLUSIONS

This paper proposes an innovative solution for broadcasting information using BLE beacons.
Field experiment shows how the system behavesto discover users and to broadcastthe information to them. However, to correctly receive the information, a userhas to install the app that can process a beacon signal. The results gave a closer look into system performance while users interact with the system. The developed system is flexible and can be incorporated easily into already built infrastructure, not ruining its scalability. The proposed system also presents an inexpensive alternative to the Cisco CMX solution, previously mentioned in the related work section.

Currently, the solution can offer a user profile categorization by applying a predefined user interest taxonomy to the system and in that way using fewer devices to advertise different data to users found in beacon proximity. That is to say, a single beacon device in our solution can operate in a mode where it associates with different advertisements and in the same moment can broadcast multiple information to nearby smart devices. Categorization process demands to predetermine the users in the ibeacon database by creating separate profiles for specific user. However, even if we do so, we have to assign them to the discovered users. That way, the system can identify them the next time they enter the beaconing area. This process asks for applying different methods for identifying users and predefining their profiles. We are currently working on a method where we try to categorize the users by analysing their movements along the building area and learning their behaviour. This way, we want to combine the dynamic broadcasting solution with the user and space analytics.

This approach enables to obtain users' movement around a specific area and, consequently, it might support discovering patterns of users' behaviour. Previously mentioned motivates as to develop and experiment more in the area of BLE beacon message dissemination. We plan to continue our work on this concept and combine our broadcasting system with space monitoring where extracting data of discovered users from beacon devices can help us to understand the surrounding behaviour, and to improve the efficiency and the logistics in the particular building. Furthermore, this approach has some numerous cases where it can be applied, such as social network, where people can search for other people around them and get some extra knowledge about the others. Also, by accessing the real-time statistic, beacons can be used to recognize different hotspots and send alerts to whom it may concern about locations that are overcrowded and a subject to security flaws.

## ACKNOWLEDGEMENT

This work is funded by: the EU Regional Development Fund, and the Galician Government under agreement for funding the (AtlantTIC); the Spanish Ministry of Economy and Competitiveness (TEC2014-54335-C4-3-R); and the EU Commission under the Erasmus Mundus Green-Tech-WB project (551984-EM-1-2014-1-ES-ERA MUNDUS-EMA21).Thanks to the Communications Department of the ATIC (UVIGO) for its collaboration in the pilot deployment

## Authors

**Miran Borić**
Erasmus Mundus Green-Tech-WB Ph.D. student since September 2015 at University of Vigo (Spain). I obtained the diploma in "Multisensor Intrusion Detection System" from the Faculty of Informational Technologies, University of "Džemal Bijedić" Mostar, Bosnia & Herzegovina. I have previous work experience as a Network specialist and System administrator at UniCredit Group Bank Mostar and Faculty of Informational Technologies Mostar. I am also an active Cisco Certified Network Academy instructor at Cisco Local Academy Mostar since 2010. My area of interest includes Smart 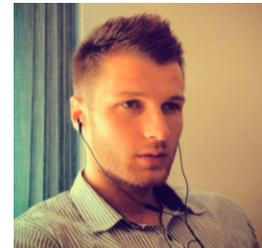 network technologies, Internet of Things, Bluetooth Technologies, Switching networks, Routing protocols and Quality of Service in computers networks.

**Ana Fernández Vilas**

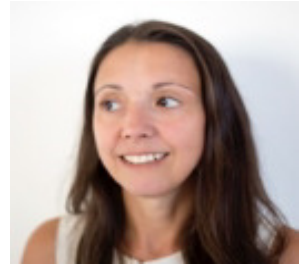

I am Associate Professor at the Department of Telematics Engineering of the University of Vigo and researcher in the Information & Computing Laboratory (AtlantTIC Research Center). I received my PhD in Computer Science from the University of Vigo in 2002. My research activity at I&C lab focuses on Semantic-Social Intelligence & data mining. I look for applying both to Ubiquitous Computing and Sensor Web; urban planning & learning analytics. Also, I am involved in several mobility & cooperation projects with North African countries & Western Balkans

**Rebeca P. Díaz Redondo**

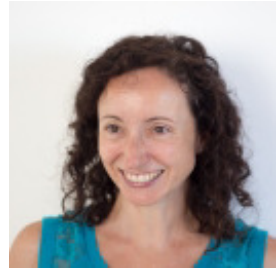

Rebeca P. Díaz Redondo is an Associate Professor at the Department of Telematics Engineering of the University of Vigo and researcher in the Information & Computing Laboratory (AtlantTIC Research Center). She received her PhD in Telecommunications Engineering from the same university. She currently works on applying social mining and data analysis techniques to characterize the behavior of users and communities to design solutions in learning, smart cities and business areas. She is currently involved in the scientific and technical activities of several national and European research & educative projects. Besides, she is involved in several mobility & cooperation projects with North African countries & Western Balkans.